# AIDA: ACCELERATOR INTEGRATED DATA ACCESS

M. Clausen (DESY), R. MacKenzie, R. Sass, H. Shoaee, K. Underwood, G. White.
Stanford Linear Accelerator Center, Stanford, CA 94309, USA

*Abstract*

All Control Systems that grow to any size have a variety of data that are stored in different formats on different nodes in the network. Examples include sensor value and status, archived sensor data, device oriented support data and relationships, message logs, application and machine configurations etc. Each type of data typically has a different programming interface. Higher-level applications need to access a logically related set of data that is in different data stores and may require different processing. AIDA is envisioned to be a distributed service that allows applications access to this wide variety of Control System data in a consistent way that is language and machine independent. It has the additional goal of providing an object-oriented layer for constructing applications on top of multiple existing conventional systems like EPICS or the SLC Control System. Motivation, design overview and current status will be presented.

## 1. MOTIVATION

At SLAC, our Control System is facing two major challenges: the uncertain long-term future of VMS and the construction of the Next Linear Collider (NLC).

When it is built, the NLC will require a Control System (NLCCS) at least an order of magnitude larger than the existing one. The requirements include that virtually all of the data, including pulsed beam data be archived all of the time. This gives the NLCCS the flavor of a detector in addition to that of a distributed Control System.

The architecture of the existing SLC Control System was designed in the early 1980s and has been providing reliable accelerator control ever since. The current primary experiment called the Asymmetric B-Factory is projected to run at least another 10 years. There are also numerous fixed-target experiments that will continue to be interspersed with the running of the B-factory. An X-ray FEL laser of unprecedented intensity is also likely to be in our future. This means that the existing Control System's reliability and availability must continue uninterrupted, but its infrastructure must be reengineered to support a migration to new platforms and architectures. The resulting Control System must be extendable and scalable for the indefinite future.

## 2. AIDA ROLE AND KEY FEATURES

The Accelerator Integrated Data Access (AIDA) system, will be the backbone of the software architecture at the middleware and higher levels. Its role is to interface application programs to subsystem's data that may be implemented in EPICS or the existing B-factory Control System. These data includes archive, configuration, model and the so-called "enterprise" data, for instance in ORACLE. It will provide the basic "get", set" and "monitor" API to all non-hard real-time data for those applications.

The objective of AIDA's architecture is to isolate the applications from dependence on the underlying data access mechanism or API. The AIDA architecture must be:

1. Distributed and highly scalable
2. Language independent. Support at least C, C++ and Java, with easy links to legacy Fortran applications.
3. Operating System independent supporting at least VMS, Win32, Unix and Linux
4. Support easy distributed development.

The fundamental solution is a course grained, so called "thin-pipe" distributed client/server 3-tier architecture based on CORBA. The logical architecture is a combination of a comprehensive name service system and a distributed 2-ply data server hierarchy as shown in Figures 1 and 2.

The location and method of data acquisition for each data item accessible by AIDA is kept in an Oracle db managed by the AIDA name service. The db also says which "Data Service" is responsible for transactions with each data item. This is similar to CDEV but with an extra level of aggregation which, for example, allows each service, such as the *magnet* service, to normalize all data into the same units regardless of which underlying subsystem the data came from, and to synchronize aggregate transactions among data sources.

For instance, say an orbit correction application wished to set the energization level of a number of dipoles simultaneously to effect a steering solution but some magnets were controlled by an EPICS control network and others were controlled by the legacy

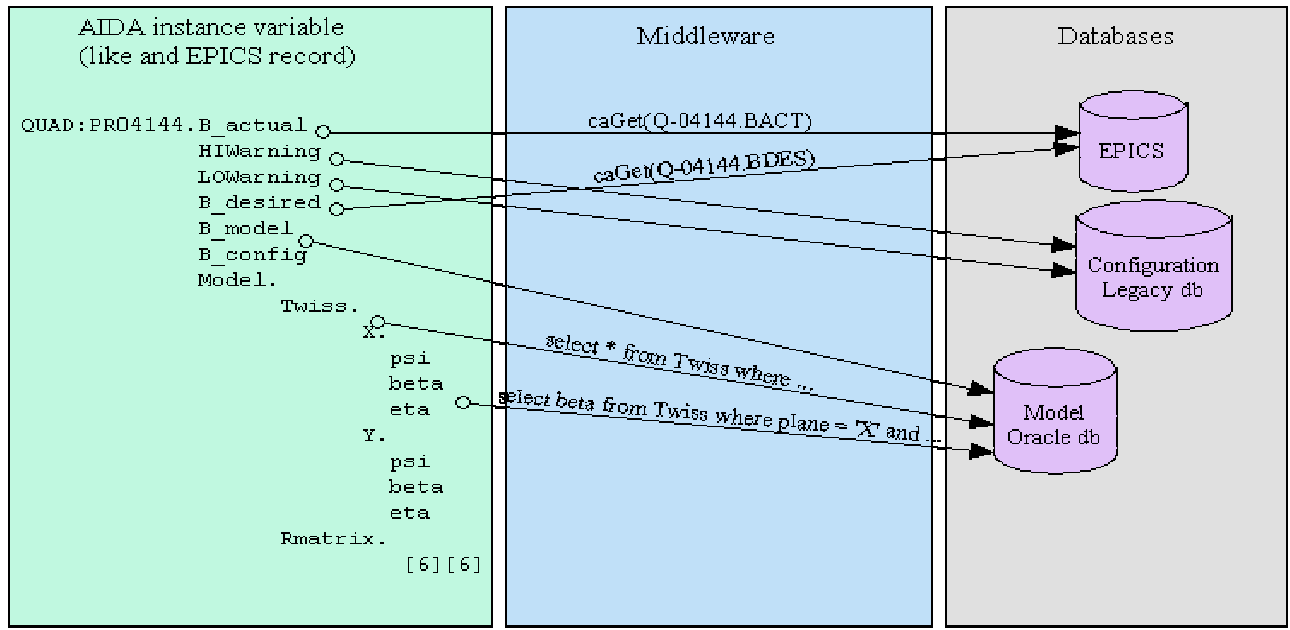

Figure 1: Logical Quadrupole device consists of data from different sources but accessible to the application via a single named hierarchy

SLC/B-factory Control System. Then the magnet service makes sure all required Δθ solutions are translated to the appropriate units for each magnet's Control System, that all magnets in each Control System are moved together and that the right synchronization method, pend_io() for EPICS and msg_send() for SLC/B-factory, is employed for each group of magnets.

As such, AIDA is not at present a comprehensive OO architecture which specifies a framework for polymorphism and encapsulation of accelerator control objects like XAL or Abeans; rather it's a high level OO API and course grained distributed client/server system. Its APIs are designed to support accelerator object libraries above it, as well as new display management frameworks and EJBs. The primary differences to CDEV are:
1. The multi-platform, multi-language support deriving from CORBA
2. The name service based on a relational database
3. Support for virtual accelerator and history data acquired through the same API as real data
4. Callback functionality based on the Corba push_consumer framework and CORBA Notify service.

A display generation and management tool started at SLAC but being developed actively by DESY, called JoiMint [1], will interface directly to AIDA.

Both procedural and object oriented APIs are planned, though at present we are concentrating on providing an OO framework. The framework includes client and server sides, so AIDA is intended as an extensible architecture to help programmers easily create new AIDA functionality.

## 2.1 The AIDA Name Service

The AIDA persistence model is that the AIDA name service uses an Oracle db to describe the method of transaction with each data source server. Each data service server, such as the magnet server, must query the name service db to discover in which data source server the named data resides, plus details of how to perform transactions on that data. For instance, the AIDA name service db record for an EPICS channel access variable will contain the pv name of the variable; the name service db record for a variable which is itself in an SQL accessible db may literally contain the SQL *select* statement or a precompiled OCI template which accesses the data. An AIDA name service record may contain one or more other AIDA record names. Hence a single Aida record may map to an n-dimensional, variable length, family of variables. Additionally the AIDA naming service allows data to be viewed and retrieved "hierarchically". For instance, one could get a structure of all the Courant Snyder parameters of a quadrupole in one call, or in fact get the history of all of these values from time a to time b, in one call as shown in see Figure 1. In this case the result is returned to the caller in a CORBA structure based on the CORBA *any*; if it is simple the structure contains only a fundamental type.

Additionally, though not yet implemented, the naming system will understand pre-defined collections, regular expressions and *range names*. Range names will allow a client to ask, for instance, for all the Twiss parameters from psi to beta say, without having to know the indexes of these variables in the structure returned by the model server.

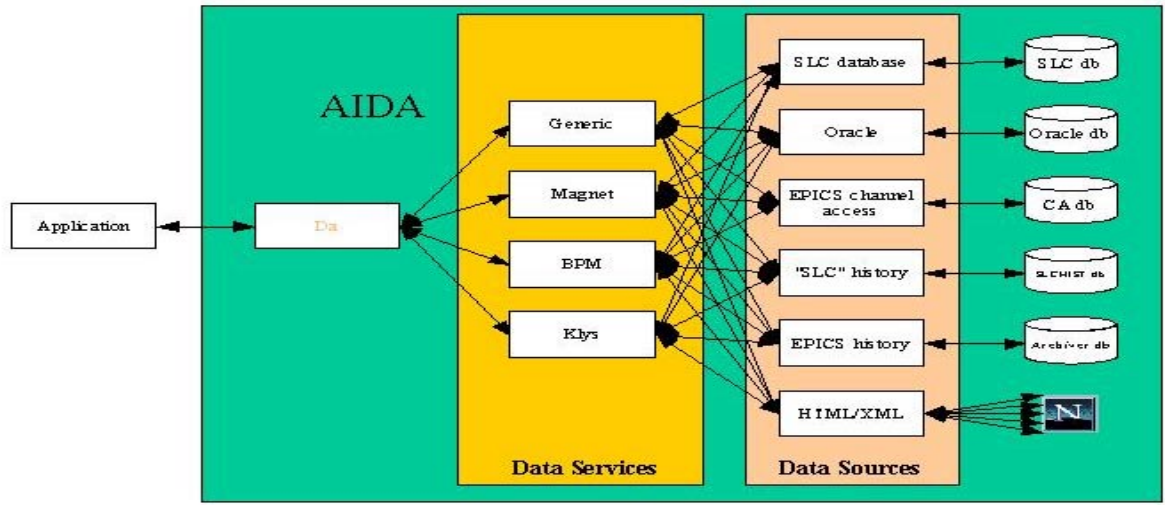

Figure 2: Logical Architecture, showing get/set/monitor, to/from any data source through a data service

The meta-data managed by the name service can persist in the client so repeated calls for the same data can go straight to the data source server that holds it.

Aida understands URL's, so data items not in the Aida name database can also be accessed through Aida's API.

## 3. MAJOR DESIGN SOLUTIONS

A course grained 3-tier client server system based on CORBA will provide the language and operating system independence as shown in Figure 2. CORBA will provide many of the required characteristics of a distributed Control System such as real-time capabilities, resource control, network quality of service, server management and asynchronous operations.

A comprehensive distributed development environment is being worked on in parallel to the AIDA functionality. This includes cvs and cvsWeb for distributed source code management; and we've found Netbeans to be a stable multi-platform (free) IDE that understands cvs natively [3]. A CORBA plug-in is available for Netbeans but we haven't tried it.

An exception handling framework that helps programmers write code, was a requirement of AIDA. In AIDA, the act of throwing an exception also issues the exception to a message client head. The framework also implements exception translation and exception tracing [2], so exceptions at all levels match the degree of abstraction of the code, but their antecedents can be traced to the root cause and the message head can be asked for the exception trace.

## 4. STATUS, PLANS AND UNKNOWNS

AIDA can now access, for get() operations, any data in the B-factory/NLCTA system, which includes the legacy VMS and new EPICS Control Systems.

The servers are presently being prototyped in Java, and so far we have had success on Unix and Windows platforms, as well as VMS! We still have some outstanding unknowns here:
1. How, if at all, Enterprise Java Beans may be used to manage data integrity and persistence
2. Whether to use an application server for server management, messaging and web interfaces
3. Which message service foundation to use (CORBA, Java, or commercial one like Tibco)
4. Whether to change the CORBA wire protocol from IIOP to some multi-cast system.

We will convert this prototype system to production quality, complete the set and monitor capability for collections and synchronized data. Then we will implement the first applications which will leverage the enterprise data capability of AIDA as part of the COSMIC migration challenge [4].